\documentclass[12pt]{iopart}
\usepackage{iopams} 

\newtheorem{proposition}{Proposition}
\newenvironment{proof}[1][Proof]{\noindent\textbf{#1.} }{\noindent \rule{0.5em}{0.5em}}

\begin{document}

\title{Reaching Fleming's discrimination bound}
\author{G Gr\"{u}bl and L Ostermann}
\address{Institut f\"{u}r Theoretische Physik der Universit\"{a}t Innsbruck\\Technikerstra\ss e 25, A-6020 Innsbruck, Austria}
\eads{\mailto{gebhard.gruebl@uibk.ac.at}, \mailto{laurin.ostermann@uibk.ac.at}}
\pacs{03.65.Ta, 03.65.Wj, 03.67.Hk}

\begin{abstract}
Any rule for identifying a quantum system's state within a set of two
non-orthogonal pure states by a single measurement is flawed. It has a
non-zero probability of either yielding the wrong result or leaving the query
undecided. This also holds if the measurement of an observable $A$ is repeated
on a finite sample of $n$ state copies. We formulate a state identification
rule for such a sample. This rule's probability of giving the wrong result
turns out to be bounded from above by $1/n\delta_{A}^{2}$ with $\delta
_{A}=\left\vert \left\langle A\right\rangle _{1}-\left\langle A\right\rangle
_{2}\right\vert /\left(  \Delta_{1}A+\Delta_{2}A\right)  .$ A larger
$\delta_{A}$ results in a smaller upper bound. Yet, according to Fleming,
$\delta_{A}$ cannot exceed $\tan\theta$ with $\theta\in\left(  0,\pi/2\right)
$ being the angle between the pure states under consideration. We demonstrate
that there exist observables $A$ which reach the bound $\tan\theta$ and we
determine all of them.
\end{abstract}

\section{Introduction}

The transmission of a binary sequence through a sequence of quantum systems
whose states are to be chosen from a given set $\left\{  \rho_{1},\rho
_{2}\right\}  $ makes it necessary for the recipient to identify the states
$\rho_{1}$ and $\rho_{2}$ with as little an error as possible. If each single
bit is transmitted as a single system, an error minimizing strategy is needed
in order to identify this system's state from one single measurement.

If it is to be discriminated between two non-orthogonal states $\rho_{1}$ and
$\rho_{2}$ through the measurement of an observable a certain \emph{positive
lower bound} for the probability of either wrong or inconclusive state
identification cannot be underrun. Such limitations of individual state
identification have been investigated extensively in theory and experiment.
For a review see e.g. \cite{Ch00}.

If in contrast each bit is transmitted as a sample of $n$ identical systems,
all of them in the same state $\rho\in\left\{  \rho_{1},\rho_{2}\right\}  , $
the minimum error in reading the message correctly reduces beyond the limit
established for $n=1.$ More generally, the sequence of values, obtained by
measuring an arbitrary, perhaps non-optimal observable $A$ on each of the
sample's members, can be used to lower the probability of an erroneous state
identification below the one obtained for a single measurement of $A.$

We will describe a rule of state identification from the mean value of a
general observable $A$ in an $n$-sample. For this rule we derive an
\emph{upper bound} for the probability of error from Chebyshev's inequality
associated with the mean value of $A.$ It turns out that our rule of state
identification produces the wrong result with a probability not greater than
$1/n\delta_{A}^{2}.$ Here the dimensionless parameter $\delta_{A}>0$ not only
depends on the two states $\rho_{1},\rho_{2}$ but also on the observable $A$
to be measured on the sample's elements.\footnote{Since we will vary $A$ and
keep the states fixed we refrain from using the more suggestive but cumbersome
notation $\delta_{A}\left(  \rho_{1},\rho_{2}\right)  .$} It is given
by\footnote{The notation seems obvious and it is spelled out in sect. 2.}
\begin{equation}
\delta_{A}=\frac{\left\vert \left\langle A\right\rangle _{\rho_{1}%
}-\left\langle A\right\rangle _{\rho_{2}}\right\vert }{\Delta_{\rho_{1}%
}A+\Delta_{\rho_{2}}A}.\label{discern}%
\end{equation}
This number therefore quantifies how well the states $\rho_{1}$ and $\rho_{2}$
can be distinguished from each other by means of measuring the observable $A.$

We will address the issue of which observable $A,$ for given states $\rho
_{1},\rho_{2},$ leads to the largest possible value of $\delta_{A}.$ Such a
choice then minimizes the upper bound $1/n\delta_{A}^{2}$ of the probability
of error for a given sample size $n,$ yet it does not need to minimize the
actual error itself. For arbitrary \emph{pure} states $\rho_{1} $ and
$\rho_{2}$ we find the maximum of $\delta_{A}$ over the set of all linear,
bounded and self-adjoint operators $A.$ We prove that
\begin{equation}
\max_{A}\delta_{A}=\tan\theta,\label{opt}%
\end{equation}
where $\theta$ with $0<\theta<\pi/2$ denotes the angle between the states
$\rho_{1}$ and $\rho_{2}.$\footnote{This means that $\tr\left(  \rho_{1}%
\rho_{2}\right)  =\cos^{2}\theta$ holds.} Furthermore, among all bounded
observables $A$ we explicitly specify those which maximize $\delta_{A}.$

The plan of the paper is as follows. In section 2 we summarize some results
concerning optimal state discrimination for single systems and exhibit their
relation to our minimization problem. In section 3 we derive the law of large
numbers which motivates our quest for maximizing $\delta_{A}.$ In section 4 we
slightly adapt Fleming's derivation of the estimate $\delta_{A}\leq\tan\theta$
to our goals. This proof will then be used first in section 5 to demonstrate
that the upper bound $\tan\theta$ can be attained and afterwards in section 6 to
identfy those observables $A$ which actually reach this bound.

\section{State identification for a single system}

How is the state $\rho$ of a single quantum system to be identified within a
given set $\left\{  \rho_{1},\rho_{2}\right\}  $ of two different yet
non-orthogonal pure states $\rho_{1}$ and $\rho_{2}?$ Is there an observable
$A$ which when measured upon $\rho$ allows for identifying the state as either
$\rho_{1}$ or $\rho_{2}$ most 'reliably'?

One way to render this question more precisely has been specified by Jaeger
and Shimony. \cite{JS95} It meanwhile bears the title \emph{'minimum error
state discrimination'}. \cite{B01}\ Assume that the spectrum of the observable
$A$ consists of two eigenvalues $a_{1},a_{2}$ only. If the system, whose state
is to be identified, is in the state $\rho_{i}$ with probability $p_{i},$ then
a measurement of a fixed observable $A$ upon a randomly chosen state yields a
random trial with the composite event space $\Omega^{A}=\left\{  \rho_{1}%
,\rho_{2}\right\}  \times\left\{  a_{1},a_{2}\right\}  $ and with the
probability measure $W^{A}$ whose distribution function $p^{A}$ obeys%
\begin{equation}
p^{A}\left(  \rho_{i},a_{j}\right)  =p_{i}\cdot \tr\left(  \rho_{i}P_{a_{j}%
}^{A}\right)  .
\end{equation}
Here $P_{a}^{A}$ denotes the orthogonal projection onto the eigenspace of $A$
corresponding to the eigenvalue $a\in\left\{  a_{1},a_{2}\right\}  .$ The
positive numbers $p_{1},p_{2}$ have to obey $p_{1}+p_{2}=1.$ In order to
correlate the measurement outcome $a_{i}$ with the random state $\rho_{i}$ as
strongly as possible one has to search for an observable $A$ which maximizes
the probability
\begin{equation}
W^{A}\left(  D\right)  =\sum_{i=1}^{2}p_{i}\cdot \tr\left(  \rho_{i}P_{a_{i}%
}^{A}\right)
\end{equation}
of the 'detection event' $D=\left\{  \left(  \rho_{1},a_{1}\right)  ,\left(
\rho_{2},a_{2}\right)  \right\}  .$

Since the states $\rho_{1},\rho_{2}$ are supposed to be pure, there exist
unitvectors $\psi_{i}\in\mathcal{H}$ such that $\rho_{i}=\psi_{i}\left\langle
\psi_{i},\cdot\right\rangle $ for $i\in\left\{  1,2\right\}  .$ Assume now
tentatively that $\tr\left(  \rho_{i}P_{a_{j}}^{A}\right)  =0$ for all pairs
$\left(  i,j\right)  $ with $i\neq j.$ This implies $P_{a_{i}}^{A}\psi_{j}=0$
for $i\neq j$ and therefore $A\psi_{i}=a_{i}\psi_{i}$ for all $i.$ But this
leads to $\left\langle \psi_{1},\psi_{2}\right\rangle =0$ which contradicts
the assumed non-orthogonality $\tr \left(  \rho_{1}\rho_{2}\right)  \neq0.$
Therefore the detection event $D$ cannot be certain whatever choice of $A$ is
made. Rather Jaeger and Shimony have proven in \cite{JS95} that the maximum of
$W^{A}\left(  D\right)  $ obeys%
\begin{equation}
\max_{A}\left\{  W^{A}\left(  D\right)  \right\}  =\frac{1}{2}\left(
1+\sqrt{1-4p_{1}p_{2}\cdot\cos^{2}\theta}\right) \label{minerrmax}%
\end{equation}
with $\theta\in\left(  0, \pi/2\right)  $ such that $\cos^{2}%
\theta=\left\vert \left\langle \psi_{1},\psi_{2}\right\rangle \right\vert
^{2}=\tr\left(  \rho_{1}\rho_{2}\right)  .$ Here the maximum is taken over all
linear, bounded and self-adjoint operators $A:\mathcal{H}\rightarrow\mathcal{H}$
whose spectrum consists of two fixed (unequal) eigenvalues $a_{1},a_{2}$ only.
It comes with little surprise that $\max_{A}\left\{  W^{A}\left(  D\right)
\right\}  $ does not depend on the choice of eigenvalues $\left(  a_{1}%
,a_{2}\right)  .$

A related maximization problem is the following one. Find a linear, bounded
and self-adjoint operator $A:\mathcal{H}\rightarrow\mathcal{H}$ such that firstly
the spectrum of $A$ consists of the eigenvalues $a_{1}=a>0,a_{2}=-a$ and
secondly $A$ maximizes the weighted difference of expectation valus, i.e.
$\Delta:=p_{1}\left\langle A\right\rangle _{1}-p_{2}\left\langle
A\right\rangle _{2}$ with $\left\langle A\right\rangle _{i}=\tr\left(  \rho
_{i}A\right)  .$ Because of
\numparts
\begin{eqnarray}
\Delta  &=ap^{A}\left(  \rho_{1},a\right)  +ap^{A}\left(  \rho_{2},-a\right)
-ap^{A}\left(  \rho_{1},-a\right)  -ap^{A}\left(  \rho_{2},a\right) \\
& =a\cdot\left(  2W^{A}\left(  D\right)  -1\right)
\label{DiffW}\
\end{eqnarray}
\endnumparts
this maximization problem for constant $a$ is equivalent to the previous one
of maximizing $W^{A}\left(  D\right)$.

A genuinely alternative maximization problem is posed by the following one,
which is known as \emph{'unambiguous state discrimination'}. \cite{B01} Assume
now that the spectrum of the observable $A$ consists of three (different)
eigenvalues $a_{0},a_{1},a_{2}.$ Then the above probability space $\left(
\Omega^{A},W^{A}\right)  $ is replaced by the event space $\Omega^{A}=\left\{
\rho_{1},\rho_{2}\right\}  \times\left\{  a_{0},a_{1},a_{2}\right\}  $ with
the modified probability measure $W^{A}$ whose distribution function $p^{A}$
obeys
\begin{equation}
p^{A}\left(  \rho_{i},a_{j}\right)  =p_{i}\cdot \tr \left(  \rho_{i}P_{a_{j}%
}^{A}\right)  .
\end{equation}
If one now chooses $A$ in such a way that
\begin{equation}
p^{A}\left(  \rho_{1},a_{2}\right)  =0=p^{A}\left(  \rho_{2},a_{1}\right)
,\label{Distinct}
\end{equation}
then, whenever the event $\left\{  \rho_{1},\rho_{2}\right\}  \times\left\{
a_{i}\right\}  $ occurs, it follows that $\rho=\rho_{i}.$ Thus, under these
provisions, the state can be determined with certainty, whenever a measurement
of $A$ yields one of the values $a_{1}$ or $a_{2}.$ Note that for $D=\left\{
\left(  \rho_{1},a_{1}\right)  ,\left(  \rho_{2},a_{2}\right)  \right\}  $
holds
\begin{equation}
W^{A}\left(  D\right)  =W^{A}\left(  \left\{  \rho_{1},\rho_{2}\right\}
\times\left\{  a_{1},a_{2}\right\}  \right)  =1-W^{A}\left(  \left\{  \rho
_{1},\rho_{2}\right\}  \times\left\{  a_{0}\right\}  \right)  .
\end{equation}
Yet, as above, the event $D=\left\{  \left(  \rho_{1},a_{1}\right)  ,\left(
\rho_{2},a_{2}\right)  \right\}  ,$ allowing for a correct
state-identification, cannot be certain. Therefore one is led to search for
those observables $A,$ for which in addition to the validity of equation
(\ref{Distinct}) the probability $W^{A}\left(  D\right)  $ is maximal.

Jaeger and Shimony \cite{JS95} have proven for $\dim\left(  \mathcal{H}\right)
\geq3$ that
\begin{equation}
\max_{A}\left\{  W^{A}\left(  D\right)  \right\}  =
\cases{
1-2\sqrt{p_{1}p_{2}}\cdot\cos\theta & for $ \sqrt{\frac{\min\left\{ p_{1},p_{2}\right\}  }{\max\left\{  p_{1},p_{2}\right\}  }} \geq \cos \theta$ \\
\max \left\{  p_{1},p_{2}\right\}  \sin^{2}\theta & for $\sqrt{\frac {\min\left\{  p_{1},p_{2}\right\}  }{\max\left\{  p_{1},p_{2}\right\}  }} <\cos\theta$ }
\label{MaxDetect}
\end{equation}
Here the maximization is performed over all those linear, bounded operators
$A,$ whose spectrum contains three different eigenvalues only, and which obey
equation (\ref{Distinct}).

In this work we shall consider a third maximization problem. Among all linear,
bounded and self-adjoint operators $A:\mathcal{H}\rightarrow\mathcal{H}$ we
determine those which maximize the number $\delta_{A}$ given by equation
(\ref{discern}) for two arbitrary but fixed non-orthogonal, non-identical, pure
state density operators $\rho_{1},\rho_{2}:\mathcal{H}\rightarrow\mathcal{H}.$
Here
\begin{equation}
\Delta_{\rho}A=\sqrt{\left\langle A^{2}\right\rangle _{\rho}-\left\langle
A\right\rangle _{\rho}^{2}}\textrm{ with }\left\langle X\right\rangle _{\rho
}=\tr\left(  \rho X\right)
\end{equation}
denotes the uncertainty of $A$ in the state $\rho.$ The number $\delta_{A}$ is
invariant under the shift $A\rightarrow A+\mu\cdot\iota d_{\mathcal{H}}$ for
any real $\mu$ and also under the rescaling $A\rightarrow\lambda A$ for any
non-zero real $\lambda.$ It relates the distance between the states'
expectation values to their uncertainties and therefore has been proposed by
Fleming \cite{F01} as a quantifier of the distinguishability of the states
$\rho_{1}$ and $\rho_{2}$ by means of measuring $A$ on a finite sample.

Part of our result is
\begin{equation}
\max_{A}\left\{  \delta_{A}\right\}  =\tan\theta,\label{qualmax}%
\end{equation}
where $A$ is allowed to run through the set of all linear, bounded,
self-adjoint operators $A:\mathcal{H}\rightarrow\mathcal{H}.$ Observe that no
further restriction on the spectrum of $A$ is imposed. In particular, the
spectrum of $A$ may include a continuous part.

Among the observables $A$ maximizing $\delta_{A}$ we shall identify one which
also maximizes $\left\langle A\right\rangle _{\rho_{1}}-\left\langle
A\right\rangle _{\rho_{2}}.$ It is given by\footnote{The spectum of $A$ is
$\left\{  1,-1,0\right\}  $ if $\dim\left(  \mathcal{H}\right)  \geq3$ and
$\left\{  1,-1\right\}  $ if $\dim\left(  \mathcal{H}\right)  =2.$} $A=\left(
\rho_{1}-\rho_{2}\right)  /\sin\theta.$ This operator therefore also maximizes
the probability of correct state identification $W\left(  D\right)  $ for
$p_{1}=p_{2}=1/2$ from equation (\ref{minerrmax}). Its value is given by%
\begin{equation}
\max_{A}\left\{  W^{A}\left(  D\right)  \right\}  =\frac{1}{2}\left(
1+\sin\theta\right)  .
\end{equation}

In deriving equation (\ref{qualmax}) we make use of a powerful estimate due to
Fleming \cite{F01}, that is conceived purely by general algebraic
deliberations. Fleming called it a \textit{' quantum master inequality'}, because he was
able to derive a host of other well known quantum theoretical facts from it.
Besides taking the orthogonality of two eigenvectors to different eigenvalues
of an observable to a more general and quantitative level, Fleming's quantum
master inequality also implies Robertson's generalized uncertainty relation
$2\Delta A\Delta B\geq\left\vert \left\langle \left[  A,B\right]
\right\rangle \right\vert .$ \cite{F07}

Fleming's quantum master inequality states that, whenever $\Delta_{\rho_{1}%
}A+\Delta_{\rho_{2}}A>0,$ then $\delta_{A}\leq\tan\theta.$ We shall first
prove that Fleming's upper bound can be reached and then identify necessary
and sufficient conditions on $A$ for $\delta_{A}=\tan\theta$ to hold.

Before entering the problem of maximizing $\delta_{A}$ we will clarify the
role of $\delta_{A}$ in identifying the state from an $n$-sample of states
$\rho\in\left\{  \rho_{1},\rho_{2}\right\}  .$ We shall do so in the more
general context of identifying a probability measure $W$ on the real line
within a set of two options $\left\{  W_{1},W_{2}\right\}  .$

\section{State identification for an $n$-sample}

Let $W$ denote a probability measure on the real line with finite expectation
value $X:=\mathbb{E}_{W}\left(  \iota d_{\mathbb{R}}\right)  $ and variance
$\Delta^{2}:=\mathbb{E}_{W}\left(  \left(  \iota d_{\mathbb{R}}-X\right)
^{2}\right)  .$ Chebyshev's inequality states that for any $t\in
\mathbb{R}_{>0}$ holds%
\begin{equation}
W\left(  \left\{  x\in\mathbb{R}:\left\vert x-X\right\vert \geq t\right\}
\right)  \leq\left(  \frac{\Delta}{t}\right)  ^{2}.\label{Chebysh1}%
\end{equation}

The product space $\mathbb{R}^{n}$ together with the product measure
$W^{n}=W\times\ldots\times W$ corresponds to the random experiment of drawing
$n$ real numbers independently and each one distributed by $W.$ The mean value
of such a sample $\left(  x_{1},\ldots x_{n}\right)  \in\mathbb{R}^{n}$ is
given by the function $m_{n}:\mathbb{R}^{n}\rightarrow\mathbb{R}$ with%
\begin{equation}
m_{n}\left(  x_{1},\ldots x_{n}\right)  =\frac{1}{n}\sum\nolimits_{i=1}%
^{n}x_{i}.
\end{equation}
For the expectation value and the variance of the random variable $m_{n}$
under the measure $W^{n}$ holds%
\begin{equation}
\mathbb{E}_{W^{n}}\left(  m_{n}\right)  =X\textrm{ and }\mathbb{V}_{W^{n}%
}\left(  m_{n}\right)  =\frac{\Delta^{2}}{n}.
\end{equation}
Application of Chebyshev's inequality to $m_{n}$ thus yields the following law
of large numbers%
\begin{equation}
W^{n}\left(  \left\{  \omega\in\mathbb{R}^{n}:\left\vert m_{n}\left(
\omega\right)  -X\right\vert \geq t\right\}  \right)  \leq\frac{1}{n}%
\cdot\left(  \frac{\Delta}{t}\right)  ^{2}\label{LargeN}%
\end{equation}
The probability that the mean value of a random sample $\omega\in
\mathbb{R}^{n}$ of the distribution $W^{n}$ deviates from the expectation
value by at least a fixed value $t>0$ converges to $0$ when $n$ goes to
$\infty.$

Now, let $W_{1}$ and $W_{2}$ denote two different probabilty measures on the
real line of the above type. Their expectation values $X_{i}$ are assumed to
be unequal and without loss of generality we may assume $X_{2}>X_{1}.$ We also
suppose that at least one of the probability measures $W_{i}$ has non-zero
variance, i.e. that $\Delta_{1}+\Delta_{2}>0.$

A sample $\omega\in\mathbb{R}^{n}$ of $n$ real numbers is supposed to be
generated by either the distribution $W_{1}^{n}$ or $W_{2}^{n}.$ From the
sample's mean value one may try to guess whether the sample has been generated
by $W_{1}^{n}$ or $W_{2}^{n}.$ To this end, observe first that as a
consequence of Chebyshev's inequality (\ref{LargeN}) we have for all
$t_{1},t_{2}\in\mathbb{R}_{>0}$%
\numparts
\begin{eqnarray}
\label{ChebN}%
W_{1}^{n}\left(  \left\{  \omega\in\mathbb{R}^{n}:m_{n}\left(  \omega\right)
\geq X_{1}+t_{1}\right\}  \right)   & \leq\frac{1}{n}\cdot\left(  \frac
{\Delta_{1}}{t_{1}}\right)  ^{2},\label{ChebN1}\\
W_{2}^{n}\left(  \left\{  \omega\in\mathbb{R}^{n}:m_{n}\left(  \omega\right)
\leq X_{2}-t_{2}\right\}  \right)   & \leq\frac{1}{n}\cdot\left(  \frac
{\Delta_{2}}{t_{2}}\right)  ^{2}.\label{ChebN2}%
\end{eqnarray}
\endnumparts
Choosing now the numbers $t_{i}$ according to

\begin{equation}
t_{i}=\Delta_{i}\cdot\delta\textrm{ with }\delta=\frac{X_{2}-X_{1}}{\Delta
_{1}+\Delta_{2}}>0
\end{equation}
the estimates (\ref{ChebN1}) and (\ref{ChebN2}) turn into
\numparts
\begin{eqnarray}
W_{1}^{n}\left(  \left\{  \omega\in\mathbb{R}^{n}:m_{n}\left(  \omega\right)
\geq X_{0}\right\}  \right)   & \leq\frac{1}{n}\cdot\left(  \frac{1}{\delta
}\right)  ^{2},\label{ChebN3}\\
W_{2}^{n}\left(  \left\{  \omega\in\mathbb{R}^{n}:m_{n}\left(  \omega\right)
\leq X_{0}\right\}  \right)   & \leq\frac{1}{n}\cdot\left(  \frac{1}{\delta
}\right)  ^{2}.\label{ChebN4}%
\end{eqnarray}
\endnumparts
Here the point
\begin{equation}
X_{0}=\frac{\Delta_{2}}{\Delta_{1}+\Delta_{2}}X_{1}+\frac{\Delta_{1}}%
{\Delta_{1}+\Delta_{2}}X_{2}\label{DivPoint}%
\end{equation}
divides the interval $\left[  X_{1},X_{2}\right]  $ into a portion of length
$\frac{\Delta_{1}}{\Delta_{1}+\Delta_{2}}\cdot\left(  X_{2}-X_{1}\right)  $ to
the left of $X_{0}$ and another one of length $\frac{\Delta_{2}}{\Delta
_{1}+\Delta_{2}}\cdot\left(  X_{2}-X_{1}\right)  $ to the right of $X_{0}.$
Observe that
\begin{equation}
X_{1}+\frac{\Delta_{1}}{\Delta_{1}+\Delta_{2}}\cdot\left(  X_{2}-X_{1}\right)
=X_{2}-\frac{\Delta_{2}}{\Delta_{1}+\Delta_{2}}\cdot\left(  X_{2}%
-X_{1}\right)  =X_{0}.
\end{equation}

Let a sample $\omega\in\mathbb{R}^{n}$ be generated with probability $p_{1}>0$
through the measure $W_{1}^{n}$ or with probability $p_{2}=1-p_{1}>0 $ through
the measure $W_{2}^{n}.$ This corresponds to the composite random trial with
event space $\Omega=\left\{  1,2\right\}  \times\mathbb{R}^{n}$ with the
product measure $W$ which obeys for all measurable $Z\subset\mathbb{R}^{n}$%
\begin{equation}
W\left(  \left\{  i\right\}  \times Z\right)  =p_{i}\cdot W_{i}^{n}\left(
Z\right)  .
\end{equation}
Let $E$ denote the event that the sample $\omega$ is either generated by
$W_{2}^{n}$ and yields a value $m_{n}\left(  \omega\right)  \leq X_{0}$ or is
generated by $W_{1}^{n}$ and has a mean value $m_{n}\left(  \omega\right)
\geq X_{0}.$ Then this event's probability is bounded by%
\begin{equation}
W\left(  E\right)  \leq p_{1}\cdot\frac{1}{n}\cdot\left(  \frac{1}{\delta
}\right)  ^{2}+p_{2}\cdot\frac{1}{n}\cdot\left(  \frac{1}{\delta}\right)
^{2}=\frac{1}{n}\cdot\left(  \frac{1}{\delta}\right)  ^{2}.
\end{equation}
Hence, by increasing $n,$ this event's probability can be made arbitrarily small.

Thus we have the result: The event that $\omega$ is generated by $W_{1}^{n}$
and has a mean value $m_{n}\left(  \omega\right)  \leq X_{0}$ or that $\omega$
is generated by $W_{2}^{n}$ and has a mean value $m_{n}\left(  \omega\right)
\geq X_{0}$ has a probabilty greater or equal to $1-\frac{1}{n}\cdot\left(
\frac{1}{\delta}\right)  ^{2}.$ For $n\delta^{2}>1/\varepsilon\gg1$ this
implies that the event $m_{n}\left(  \omega\right)  \leq X_{0}$ is caused by
$W_{1}^{n}$ and $m_{n}\left(  \omega\right)  \geq X_{0}$ is caused by
$W_{2}^{n}$ has probability greater than $1-\varepsilon$ and so is virtually
certain. This fact justifies the identification of a sample's generating
distribution by means of the following criterion: if the sample's mean value
obeys $m_{n}\left(  \omega\right)  <X_{0},$ then the sample $\omega$ is
assumed to have been generated by $W_{1}^{n}.$ If, however, $m_{n}\left(
\omega\right)  \geq X_{0},$ then the sample is assumed to have been gererated
by $W_{2}^{n}.$

Besides the sample's size $n$ the positiv real number
\begin{equation}
\delta=\frac{\left\vert X_{1}-X_{2}\right\vert }{\Delta_{1}+\Delta_{2}%
}\label{UbD}%
\end{equation}
is decisive for the correct identification of the sample-generating
distribution $W_{i}^{n}$ from the sample's value $m_{n}\left(  \omega\right)
$ with high probability. The two distributions $W_{1}^{n}$ and $W_{2}^{n}$ are
identified correctly with the higher a probability the larger the value of
$\delta.$ One might call the parameter $\delta$ of two probability measures
$W_{1}$ and $W_{2}$ on the real line their \emph{discernability.}

As is well known, any pair $\left(  \rho,A\right)  $ of a quantum state
$\rho:\mathcal{H}\rightarrow\mathcal{H}$ and a bounded observable
$A:\mathcal{H}\rightarrow\mathcal{H}$ defines a probability measure $W_{\rho
}^{A}$ on $\mathbb{R},$ which has its support on the spectrum of $A.$ For any
measurable set $Z\subset\mathbb{R}$ the number $W_{\rho}^{A}\left(  Z\right)
$ equals the probability that when $A$ is measured on $\rho$ the measured
value belongs to $Z.$ The expectation value and variance of $\iota
d_{\mathbb{R}}$ under $W_{\rho}^{A}$ equal $\tr\left(  \rho A\right)
=\left\langle A\right\rangle _{\rho}$ and $\tr\left(  \rho A^{2}\right)
-\tr\left(  \rho A\right)  ^{2}=\left(  \Delta_{\rho}A\right)  ^{2}.$ Thus the
rule of identifying a probability measure $W\in\left\{  W_{1},W_{2}\right\}  $
from an $n$-sample of measured values $\left(  \omega_{1},\ldots\omega
_{n}\right)  $ can be taken over in a straight-forward manner to the quantum
case by replacing $W_{i}$ with the probability measure $W_{\rho_{i}}^{A}.$
Identification of $W_{i}$ then amounts to an identification of $\rho_{i}$ and
the discernibility $\delta$ specializes to the expression given by equation
(\ref{discern}).

\section{Fleming's quantum master inequality}

Let $\mathcal{H}$ denote a separable Hilbert space and let the linear mapping
$A:\mathcal{H}\rightarrow\mathcal{H}$ be bounded and self-adjoint. The
expectation value of $A$ in the pure state represented by a unit vector
$v\in\mathcal{H}$ is denoted as $\left\langle A\right\rangle _{v}%
.$\footnote{Thus $\left\langle A\right\rangle _{v}=\left\langle
v,Av\right\rangle ,$ and $\Delta_{v}A=\sqrt{\left\langle A^{2}\right\rangle
_{v}-\left\langle A\right\rangle _{v}^{2}}$ denotes the uncertainty of $A$ in
the state represented by $v.$} The following \emph{quantum master inequality}
(QMIE) relates two pure states through their first two moments of an
observable. It has been given by Fleming in \cite{F01}.

\begin{proposition}
\label{Flem1}For any two unit vectors $v,w\in\mathcal{H}$ and any linear,
bounded and self-adjoint operator $A:\mathcal{H}\rightarrow\mathcal{H}$ there
holds%
\begin{equation}
\left\vert \left\langle A\right\rangle _{w}-\left\langle A\right\rangle
_{v}\right\vert \cdot\left\vert \left\langle w,v\right\rangle \right\vert
\leq\left(  \Delta_{v}A+\Delta_{w}A\right)  \cdot\sqrt{1-\left\vert
\left\langle w,v\right\rangle \right\vert ^{2}}.\label{estimate1}%
\end{equation}
\end{proposition}

Before proving this estimate we discuss a few of its consequences. Observe
first that there exists a unique $\theta\in\left[  0,\pi/2\right]  $ such that
$\left\vert \left\langle w,v\right\rangle \right\vert =\cos\theta.$ Squaring
the inequality (\ref{estimate1}) and slightly rearranging terms yields%
\begin{equation}
\left[  \left(  \left\langle A\right\rangle _{w}-\left\langle A\right\rangle
_{v}\right)  ^{2}+\left(  \Delta_{v}A+\Delta_{w}A\right)  ^{2}\right]
\cos^{2}\theta\leq\left(  \Delta_{v}A+\Delta_{w}A\right)  ^{2}.
\end{equation}
Whenever the term in the square brackets is non-zero, then (\ref{estimate1})
is equivalent to%
\begin{equation}
\cos^{2}\theta\leq\frac{\left(  \Delta_{v}A+\Delta_{w}A\right)  ^{2}%
}{\left\vert \left\langle A\right\rangle _{w}-\left\langle A\right\rangle
_{v}\right\vert ^{2}+\left(  \Delta_{v}A+\Delta_{w}A\right)  ^{2}%
}.\label{dist1}%
\end{equation}
For $\Delta_{v}A+\Delta_{w}A>0$ inequality (\ref{dist1}) is equivalent to%
\begin{equation}
\cos^{2}\theta\leq\frac{1}{1+\delta^{2}}\textrm{ with }\delta=\frac{\left\vert
\left\langle A\right\rangle _{w}-\left\langle A\right\rangle _{v}\right\vert
}{\Delta_{v}A+\Delta_{w}A}\geq0.\label{dist}%
\end{equation}

The bound (\ref{dist}) for $\cos^{2}\theta$ is strictly monotonically
decreasing from $1$ to $0$ when $\delta$ moves from $0$ to $\infty.$ The
number $\delta$ quantifies the distinguishability of the states represented by
$v$ and $w$ through measuring $A.$ For non-orthogonal vectors $v$ and $w$ the
inequality (\ref{dist}) is equivalent to%
\begin{equation}
\delta^{2}\leq\frac{1}{\cos^{2}\theta}-1=\tan^{2}\theta.
\end{equation}
Thus, for $\cos\theta>0$ and $\Delta_{v}A+\Delta_{w}A>0$ the estimate
(\ref{estimate1}) is equivalent to%
\begin{equation}
\delta\leq\tan\theta.\label{DeltaBound}%
\end{equation}

If $v$ and $w$ are parallel, then both sides of the inequality
(\ref{estimate1}) take the value $0$ due to $\left\langle A\right\rangle
_{w}=\left\langle A\right\rangle _{v}$ and $\left\vert \left\langle
w,v\right\rangle \right\vert =1.$ The inequality (\ref{estimate1}) is thus
saturated in this case for any $A.$ If $v$ and $w$ are orthogonal, the
estimate (\ref{estimate1}) reduces to%
\begin{equation}
0\leq\Delta_{v}A+\Delta_{w}A.
\end{equation}
This estimate is saturated if and only if $\Delta_{v}A=0=\Delta_{w}A,$ which
in turn holds if and only if both $v$ and $w$ are eigenvectors of $A.$

We shall now give a proof of Fleming's quantum master inequality
(\ref{estimate1}).

\begin{proof}
In a first step we decompose the vector $Av$ into a vector parallel to $v$ and
one orthogonal to $v.$ This unique decomposition reads%
\begin{equation}
Av=\left\langle A\right\rangle _{v}v+\left(  Av-\left\langle A\right\rangle
_{v}v\right)  ,
\end{equation}
since $\left\langle v,Av-\left\langle A\right\rangle _{v}v\right\rangle =0.$
Observe that the component $v_{A}=Av-\left\langle A\right\rangle _{v}v$ of
$Av$ orthogonal to $v$ has the norm $\Delta_{v}A$ since%
\begin{equation}
\left\vert v_{A}\right\vert ^{2}=\left\langle Av-\left\langle A\right\rangle
_{v}v,Av-\left\langle A\right\rangle _{v}v\right\rangle =\left(  \Delta
_{v}A\right)  ^{2}.
\end{equation}

We thus have
\begin{equation}
\left\langle w,Av\right\rangle =\left\langle A\right\rangle _{v}\left\langle
w,v\right\rangle +\left\langle w,v_{A}\right\rangle .
\end{equation}
The analogous decomposition of $Aw=\left\langle A\right\rangle _{w}w+w_{A}$
with $w_{A}=Aw-\left\langle A\right\rangle _{w}w$ yields%
\begin{equation}
\left\langle Aw,v\right\rangle =\left\langle A\right\rangle _{w}\left\langle
w,v\right\rangle +\left\langle w_{A},v\right\rangle .
\end{equation}
Since $\left\langle w,Av\right\rangle =\left\langle Aw,v\right\rangle $ we
obtain%
\begin{equation}
\left(  \left\langle A\right\rangle _{w}-\left\langle A\right\rangle
_{v}\right)  \left\langle w,v\right\rangle =\left\langle w,v_{A}\right\rangle
-\left\langle w_{A},v\right\rangle .
\end{equation}
Taking the absolute value from both sides and applying the triangle inequality
on $\mathbb{C}$ gives the estimate%
\begin{equation}
\left\vert \left\langle A\right\rangle _{w}-\left\langle A\right\rangle
_{v}\right\vert \cdot\cos\theta=\left\vert \left\langle w,v_{A}\right\rangle
-\left\langle w_{A},v\right\rangle \right\vert \leq\left\vert \left\langle
w,v_{A}\right\rangle \right\vert +\left\vert \left\langle w_{A},v\right\rangle
\right\vert ,\label{estimate2}%
\end{equation}
where $\theta\in\left[  0,\pi/2\right]  $ is uniquely defined through
$\cos\theta=\left\vert \left\langle w,v\right\rangle \right\vert .$

Since $v_{A}$ is orthogonal to $v,$ by means of the decomposition of $w$ into
a vector from $\mathbb{C}\cdot v$ and one from its orthogonal complement
$\left(  \mathbb{C}\cdot v\right)  ^{\bot}$ according to%
\begin{equation}
w=\left\langle v,w\right\rangle v+\left(  w-\left\langle v,w\right\rangle
v\right)  ,
\end{equation}
we obtain the equality%
\begin{equation}
\left\langle w,v_{A}\right\rangle =\left\langle w-\left\langle
v,w\right\rangle v,v_{A}\right\rangle .
\end{equation}
After taking the absolute value from both sides the Cauchy-Schwarz inequality
in $\mathcal{H}$ gives%
\begin{equation}
\left\vert \left\langle w,v_{A}\right\rangle \right\vert =\left\vert
\left\langle w-\left\langle v,w\right\rangle v,v_{A}\right\rangle \right\vert
\leq\left\vert w-\left\langle v,w\right\rangle v\right\vert \left\vert
v_{A}\right\vert =\sin\theta\cdot\Delta_{v}A.\label{estimate3}%
\end{equation}
Interchanging $v$ and $w$ leaves us with%
\begin{equation}
\left\vert \left\langle w_{A},v\right\rangle \right\vert =\left\vert
\left\langle w_{A},v-\left\langle w,v\right\rangle w\right\rangle \right\vert
\leq\sin\theta\cdot\Delta_{w}A.\label{estimate4}%
\end{equation}

Inserting these bounds into the right-hand side of estimate (\ref{estimate2})
then leads to the statement of prop. \ref{Flem1},%
\begin{equation}
\left\vert \left\langle A\right\rangle _{w}-\left\langle A\right\rangle
_{v}\right\vert \cdot\cos\theta\leq\left(  \Delta_{v}A+\Delta_{w}A\right)
\cdot\sin\theta.
\end{equation}
\end{proof}

\section{Conditions for saturating the QMIE}

When $H=\hbar h$ denotes a Hamiltonian, the
estimate (\ref{dist1}) with $w=v_{t}:=e^{-iht}v$ produces an upper bound for
the survival probability $P_{v}\left(  t\right)  =\left\vert \left\langle
v,v_{t}\right\rangle \right\vert ^{2}$ of the initial state $v\left\langle
v,\cdot\right\rangle $ under the time evolution up to time $t,$ which is
equivalent to%
\begin{equation}
0\leq Q\left(  t\right)  :=\frac{\left(  \Delta_{v}A+\Delta_{v_{t}}A\right)
^{2}}{\left(  \Delta_{v}A+\Delta_{v_{t}}A\right)  ^{2}+\left\vert \left\langle
A\right\rangle _{v_{t}}-\left\langle A\right\rangle _{v}\right\vert ^{2}%
}-P_{v}\left(  t\right)  .\label{surv}%
\end{equation}
Searching for an observable $A$ which minimizes $\int_{0}^{T}Q\left(
t\right)  dt$ for a period of time $T$ of a spin-1/2-system, we have found the
following necessary and sufficient conditions on $A$ to saturate the QMIE.
\cite{Os10}

\begin{proposition}
\label{Sat}Let $v,w$ be unit vectors in a Hilbert space $\mathcal{H}$ with
$\left\vert \left\langle w,v\right\rangle \right\vert =\cos\theta$ for some
$\theta\in\left(  0,\pi/2\right]  ,$ i.e. $v$ and $w$ are assumed to be linearly
independent. Let $A:\mathcal{H}\rightarrow\mathcal{H}$ be a linear, bounded
and self-adjoint. Then, the QMIE for the states $\rho_{1}=v\left\langle
v,\cdot\right\rangle $ and $\rho_{2}=w\left\langle w,\cdot\right\rangle $ is
saturated, i.e, the equation%
\begin{equation}
\left\vert \left\langle A\right\rangle _{w}-\left\langle A\right\rangle
_{v}\right\vert \cos\theta=\left(  \Delta_{v}A+\Delta_{w}A\right)  \sin
\theta\label{QME}%
\end{equation}
holds, if and only if the conditions (i) and (ii) are fulfilled.

\begin{enumerate}
\item The operator $A$ leaves the subspace $\mathbb{C}\cdot v+\mathbb{C}\cdot
w$ invariant.

\item The equation%
\begin{equation}
\left\langle w,Av\right\rangle =\lambda\left\langle w,v\right\rangle
\label{realprop}%
\end{equation}
holds for some $\lambda\in\mathbb{R}$ with%
\begin{equation}
\min\left\{  \left\langle A\right\rangle _{w},\left\langle A\right\rangle
_{v}\right\}  \leq\lambda\leq\max\left\{  \left\langle A\right\rangle
_{w},\left\langle A\right\rangle _{v}\right\}  .
\end{equation}

\end{enumerate}
\end{proposition}

For $\theta=0$ the QMIE is saturated for any observable $A$ since both sides
of the QMIE are zero. Thus, the proposition has to deal with the non-trivial
case $0<\theta\leq\pi/2$ only.

\begin{proof}
The proof of proposition \ref{Flem1} contains three estimates. The first one
is (\ref{estimate2}). It uses the triangle inequality for complex numbers as
follows%
\begin{equation}
\left\vert \left\langle w,v_{A}\right\rangle -\left\langle w_{A}%
,v\right\rangle \right\vert \leq\left\vert \left\langle w,v_{A}\right\rangle
\right\vert +\left\vert \left\langle w_{A},v\right\rangle \right\vert .
\end{equation}
Here, equality holds if and only if the complex numbers $\left\langle
w,v_{A}\right\rangle $ and $-\left\langle w_{A},v\right\rangle $ as elements
of $\mathbb{R}^{2}\simeq\mathbb{C}$\ point into the same direction. This is
the case if and only if there exists a pair $\left(  \alpha,\beta\right)
\in\left(  \mathbb{R}_{\geq0}\times\mathbb{R}_{\geq0}\right)  \smallsetminus
\left(  0,0\right)  $ such that%
\begin{equation}
\alpha\left\langle w,v_{A}\right\rangle +\beta\left\langle w_{A}%
,v\right\rangle =0.\label{C1}%
\end{equation}

The other two estimates are contained in (\ref{estimate3}) and
(\ref{estimate4}). They employ the Cauchy-Schwarz inequality for the scalar
product of two elements of $\mathcal{H}.$ The estimate (\ref{estimate3}) thus
is saturated if and only if the vector $v_{A}$ is a (complex) multiple of the
(non-zero)\footnote{Due to $\theta>0$ we have $w-\left\langle v,w\right\rangle
v\neq0\neq v-\left\langle w,v\right\rangle w.$} vector $w-\left\langle
v,w\right\rangle v,$ i.e., if
\begin{equation}
v_{A}\in\mathbb{C}\cdot\left(  w-\left\langle v,w\right\rangle v\right)
.\label{C2}%
\end{equation}
Similarly, the estimate (\ref{estimate4}) is saturated if and only if
\begin{equation}
w_{A}\in\mathbb{C}\cdot\left(  v-\left\langle w,v\right\rangle w\right)
.\label{C3}%
\end{equation}

Therefore, the equality (\ref{QME}) holds if and only if all three conditions
(\ref{C1}), (\ref{C2}), and (\ref{C3}) are fullfilled. The conditions
(\ref{C2}), and (\ref{C3}) hold, if and only if $A$ maps the space which is
spanned by $v$ and $w$ onto itself. This can be seen as follows: (\ref{C2})
implies\footnote{Observe that it is here that we need that $v$ and $w$ are
linearly independent, i.e. that $\theta>0.$ In case of $\theta=0$ a condition on
$v_{A}$ does not follow from saturating estimate (\ref{estimate3}).} that
$Av\in\mathbb{C}\cdot v+\mathbb{C}\cdot w$ and (\ref{C3}) implies that
$Aw\in\mathbb{C}\cdot v+\mathbb{C}\cdot w.$ On the other hand, since $v_{A}$
is by definition orthogonal to $v$ and $w_{A}$ is orthogonal to $w,$ the
conditions (\ref{C2}) and (\ref{C3}) follow from $A\left(  \mathbb{C}\cdot
v+\mathbb{C}\cdot w\right)  \subset\left(  \mathbb{C}\cdot v+\mathbb{C}\cdot
w\right)  .$

We now have to address condition (\ref{C1}). This condition says that there
exists a pair $\left(  \alpha,\beta\right)  \in\left(  \mathbb{R}_{\geq
0}\times\mathbb{R}_{\geq0}\right)  \smallsetminus\left(  0,0\right)  $ such
that
\numparts
\begin{eqnarray}
0  & =\alpha\left\langle w,Av-\left\langle A\right\rangle _{v}v\right\rangle
+\beta\left\langle Aw-\left\langle A\right\rangle _{w}w,v\right\rangle \\
& =\alpha\left\langle w,Av\right\rangle +\beta\left\langle Aw,v\right\rangle
-\alpha\left\langle A\right\rangle _{v}\left\langle w,v\right\rangle
-\beta\left\langle A\right\rangle _{w}\left\langle w,v\right\rangle \\
& =\left(  \alpha+\beta\right)  \left\langle w,Av\right\rangle -\left(
\alpha\left\langle A\right\rangle _{v}+\beta\left\langle A\right\rangle
_{w}\right)  \left\langle w,v\right\rangle .
\end{eqnarray}
\endnumparts
Since $\alpha+\beta>0,$ it follows that condition (\ref{C1}) holds if and only
if there exists a pair $\left(  \alpha,\beta\right)  \in\left(  \mathbb{R}%
_{\geq0}\times\mathbb{R}_{\geq0}\right)  \smallsetminus\left(  0,0\right)  $
such that
\begin{equation}
\left\langle w,Av\right\rangle =\left(  \frac{\alpha}{\alpha+\beta
}\left\langle A\right\rangle _{v}+\frac{\beta}{\alpha+\beta}\left\langle
A\right\rangle _{w}\right)  \left\langle w,v\right\rangle .
\end{equation}
Due to $\left(  \frac{\alpha}{\alpha+\beta},\frac{\beta}{\alpha+\beta}\right)
\in\left(  \left[  0,1\right]  \times\left[  0,1\right]  \right)
\smallsetminus\left(  0,0\right)  ,$ and $\frac{\alpha}{\alpha+\beta}%
+\frac{\beta}{\alpha+\beta}=1,$ the real number%
\begin{equation}
\lambda=\left(  \frac{\alpha}{\alpha+\beta}\left\langle A\right\rangle
_{v}+\frac{\beta}{\alpha+\beta}\left\langle A\right\rangle _{w}\right)
\end{equation}
is a convex combination of $\left\langle A\right\rangle _{v}$ and
$\left\langle A\right\rangle _{w}.$ Thus condition (\ref{C1}) holds if and
only if $\left\langle w,Av\right\rangle $ is a real multiple of $\left\langle
w,v\right\rangle ,$ where the factor belongs to the interval bounded by
$\left\langle A\right\rangle _{v}$ and $\left\langle A\right\rangle _{w}.$
Thus we have proven equation (\ref{realprop}).
\end{proof}

Observe that in case of $\cos\theta=0$ the pair $\left(  v,w\right)  $ is an
orthonormal basis of the space $\mathbb{C}\cdot v+\mathbb{C}\cdot w.$ Then,
the equation (\ref{QME}) holds if and only if $A$ stabilizes the subspace
$\mathbb{C}\cdot v+\mathbb{C}\cdot w$ and $\left\langle w,Av\right\rangle =0.$
This in turn is equivalent to the statement that $v$ and $w$ both are
eigenvectors of $A,$ because of $Av=\left\langle v,Av\right\rangle
v+\left\langle w,Av\right\rangle w=\left\langle A\right\rangle _{v}v$ and
similarly $Aw=\left\langle A\right\rangle _{w}w.$ If on the other hand for
$0<\theta<\pi/2$ we have $\Delta_{v}A+\Delta_{w}A=0$ it follows that
$\left\langle A\right\rangle _{w}=\left\langle A\right\rangle _{v}$ and that
$v$ and $w$ are eigenvectors of $A$ with the same eigenvalue.

Thus the nontrivial case of equation (\ref{QME}) is realized if $0<\theta
<\pi/2$ and $\Delta_{v}A+\Delta_{w}A>0$ is valid. In this case the equality
$\delta_{A}=\tan\theta$ holds if and only if the conditions (i) and (ii) are fulfilled.

\section{Observables of maximal $\delta_{A}$}

We shall now determine the set of observables $A$ which for given states
$\rho_{1}=v\left\langle v,\cdot\right\rangle ,$ and $\rho_{2}=w\left\langle
w,\cdot\right\rangle $ obey $\Delta_{w}A+\Delta_{v}A>0$ and $\delta_{A}%
=\tan\theta.$ Let $v,w\in\mathcal{H}$ be unit vectors with $0<\left\vert
\left\langle v,w\right\rangle \right\vert <1$ and let $A:\mathcal{H}%
\rightarrow\mathcal{H}$ be linear, bounded and self-adjoint. Without loss of
generality we assume that $\left\langle A\right\rangle _{v}\leq\left\langle
A\right\rangle _{w}$ and that $\left\vert \left\langle v,w\right\rangle
\right\vert =\left\langle v,w\right\rangle .$ According to prop. \ref{Sat} the
equation $\delta_{A}=\tan\theta$ holds if and only if

\begin{enumerate}
\item $A$ stabilizes $\mathbb{C}\cdot v+\mathbb{C}\cdot w$

\item The quotient $\frac{\left\langle w,Av\right\rangle }{\left\langle
w,v\right\rangle }$ is real and obeys $\left\langle A\right\rangle _{v}%
\leq\frac{\left\langle w,Av\right\rangle }{\left\langle w,v\right\rangle }%
\leq\left\langle A\right\rangle _{w}.$
\end{enumerate}

Since $A$ is self-adjoint, condition (i) implies that $A$ stabilizes the
orthogonal complement of $\mathbb{C}\cdot v+\mathbb{C}\cdot w$ too. Therefore,
the action of $A$ on this complementary subspace $\left[  \mathbb{C}\cdot
v+\mathbb{C}\cdot w\right]  ^{\bot}$ has no relevance to our problem and it is
the restriction $A_{0}$ of $A$ to $\mathcal{H}_{0}:=\mathbb{C}\cdot
v+\mathbb{C}\cdot w$ only which has to be studied.

Since $\delta_{\lambda A+\mu\iota d_{\mathcal{H}}}=\delta_{A}$ holds for all
$\lambda\in\mathbb{R}\setminus0$ and $\mu\in\mathbb{R}$ and for all $A$ with
$\Delta_{w}A+\Delta_{v}A>0,$ we may use this freedom of shifting and rescaling
$A$ in such a way that the spectrum of $A_{0}$ obeys $\sigma\left(
A_{0}\right)  =\left\{  1,-1\right\}  .$ This is clearly equivalent to
\begin{equation}
\tr\left(  A_{0}\right)  =0\textrm{ and }det\left(  A_{0}\right)
=-1.\label{Normal}%
\end{equation}
Observe that, because of $0<\Delta_{w}A+\Delta_{v}A=\Delta_{w}A_{0}+\Delta
_{v}A_{0},$ a transformation into $A_{0}=\iota d_{\mathcal{H}_{0}}$ is impossible.

Among the observables $A_{0}:\mathcal{H}_{0}\rightarrow\mathcal{H}_{0}$ wich
obey (\ref{Normal}) we now search for those which satisfy%
\begin{equation}
\left\langle A_{0}\right\rangle _{v}\leq\frac{\left\langle w,A_{0}%
v\right\rangle }{\left\langle w,v\right\rangle }\leq\left\langle
A_{0}\right\rangle _{w}.\label{E1}%
\end{equation}
To do so we introduce the following orthonormal basis in $\mathcal{H}_{0}:$%
\begin{equation}
e_{1}=\frac{v+w}{2\cos\left(  \frac{\theta}{2}\right)  },\quad e_{2}%
=\frac{w-v}{2\sin\left(  \frac{\theta}{2}\right)  }.\label{ONB}%
\end{equation}
The vectors $v$ and $w$ thus have the decomposition%
\begin{equation}
w=\cos\left(  \frac{\theta}{2}\right)  \cdot e_{1}+\sin\left(  \frac{\theta
}{2}\right)  \cdot e_{2}\textrm{ and }v=\cos\left(  \frac{\theta}{2}\right)
\cdot e_{1}-\sin\left(  \frac{\theta}{2}\right)  \cdot e_{2}.\label{Decomp}%
\end{equation}

The matrix elements of $A_{0}$ with respect to $\underline{e}=\left(
e_{1},e_{2}\right)  $ are denoted as $A_{ij}=\left\langle e_{i},A_{0}%
e_{j}\right\rangle .$ Clearly, $A_{ii}\in\mathbb{R}$ and $A_{12}\in\mathbb{C}$
with $A_{21}=\overline{A_{12}}.$ Condition (\ref{Normal}) is equivalent to%
\begin{equation}
A_{11}=-A_{22}\textrm{ and }A_{11}^{2}+\left\vert A_{12}\right\vert ^{2}=1.
\end{equation}

For the matrix elements involved in (\ref{E1}) we find
\numparts
\begin{eqnarray}
\left\langle v,Av\right\rangle  & =\cos\left(  \theta\right)  A_{11}%
-\sin\left(  \theta\right)  \Re\left(  A_{12}\right)  ,\\
\left\langle w,Av\right\rangle  & =A_{11}-i\sin\left(  \theta\right)
\Im\left(  A_{12}\right)  ,\\
\left\langle w,Aw\right\rangle  & =\cos\left(  \theta\right)  A_{11}%
+\sin\left(  \theta\right)  \Re\left(  A_{12}\right)  .
\end{eqnarray}
\endnumparts
Condition (\ref{E1}) therefore implies that
\[
\Im\left(  A_{12}\right)  =0\textrm{ and }\Re\left(  A_{12}\right)  \geq0.
\]
Due to $A_{11}^{2}+A_{12}^{2}=1$ there exists a unique $\alpha\in\left[
0,\pi\right]  $ with
\[
A_{11}=\cos\alpha\textrm{ and }A_{12}=\sin\alpha.
\]

Using this parametrization the matrix elements of $A_{0}$ obey
\numparts
\begin{eqnarray}
\left\langle v,Av\right\rangle  & =\cos\left(  \theta+\alpha\right)
,\label{me1}\\
\left\langle w,Av\right\rangle  & =\cos\alpha,\label{me2}\\
\left\langle w,Aw\right\rangle  & =\cos\left(  \theta-\alpha\right)
\label{me3}
\end{eqnarray}
\endnumparts
Condition (\ref{E1}) thus implies
\begin{equation}
\cos\left(  \theta\right)  \cos\left(  \theta+\alpha\right)  \leq\cos
\alpha\leq\cos\left(  \theta\right)  \cos\left(  \theta-\alpha\right)  .
\end{equation}
which, due to $\cos\left(  \theta\right)  \cos\left(  \theta+\alpha\right)
=\left[  \cos\left(  \alpha\right)  +\cos\left(  2\theta+\alpha\right)
\right]  /2,$ is equivalent to%
\begin{equation}
\cos\left(  2\theta+\alpha\right)  \leq\cos\alpha\leq\cos\left(
2\theta-\alpha\right)  .\label{E2}%
\end{equation}
On the domain $\left(  \theta,\alpha\right)  \in\left(  0,\pi/2\right)
\times\left[  0,\pi\right]  $ condition (\ref{E2}) is equivalent to%
\begin{equation}
\theta\leq\alpha\leq\pi-\theta.
\end{equation}

Finally, it is now easy to prove that for any linear, bounded and self-adjoint
map $A:\mathcal{H}\rightarrow\mathcal{H}$ which stabilizes $\mathcal{H}_{0}$
and whose restriction $A_{0}$ to $\mathcal{H}_{0}$ obeys
\begin{equation}
A_{0}=\cos\left(  \alpha\right)  \left[  e_{1}\left\langle e_{1}%
,\cdot\right\rangle -e_{2}\left\langle e_{2},\cdot\right\rangle \right]
+\sin\left(  \alpha\right)  \left[  e_{1}\left\langle e_{2},\cdot\right\rangle
+e_{2}\left\langle e_{1},\cdot\right\rangle \right] \label{Satbound}%
\end{equation}
for some $\alpha\in\left[  \theta,\pi-\theta\right]  $ there holds $\delta
_{A}=\tan\theta.$ To do so we first derive from (\ref{me1}) and (\ref{me3})%
\begin{equation}
\left\langle w,Aw\right\rangle -\left\langle v,Av\right\rangle =2\sin\left(
\theta\right)  \sin\left(  \alpha\right) \label{DiffExpt}%
\end{equation}
and then observe that
\begin{eqnarray}
\left(  \Delta_{v}A\right)  ^{2}  & =1-\left\langle A\right\rangle _{v}%
^{2}=\sin^{2}\left(  \theta+\alpha\right)  ,\\
\left(  \Delta_{w}A\right)  ^{2}  & =1-\left\langle A\right\rangle _{w}%
^{2}=\sin^{2}\left(  \theta-\alpha\right) .
\end{eqnarray}
From this it follows that%
\begin{equation}
\delta_{A}=\frac{\left\langle w,Aw\right\rangle -\left\langle
v,Av\right\rangle }{\Delta_{w}A+\Delta_{v}A}=\frac{2\sin\left(  \theta\right)
\sin\left(  \alpha\right)  }{\sqrt{\sin^{2}\left(  \theta-\alpha\right)
}+\sqrt{\sin^{2}\left(  \theta+\alpha\right)  }}.
\end{equation}
Since $0\leq\alpha-\theta\leq\pi$ and $0<\theta+\alpha\leq\pi$ we have%
\begin{equation}
\sqrt{\sin^{2}\left(  \theta-\alpha\right)  }=\sin\left(  \alpha
-\theta\right)  \textrm{ and }\sqrt{\sin^{2}\left(  \theta+\alpha\right)  }%
=\sin\left(  \alpha+\theta\right)
\end{equation}
and therefore%
\begin{equation}
\delta_{A}=\frac{2\sin\left(  \theta\right)  \sin\left(  \alpha\right)
}{2\sin\left(  \alpha\right)  \cos\left(  \theta\right)  }=\tan\left(
\theta\right)  .
\end{equation}

We may now summarize our result as follows.

\begin{proposition}
Let $v,w$ be unit vectors in a Hilbert space $\mathcal{H}$ with $\left\langle
v,w\right\rangle =\cos\theta$ for some $\theta\in\left(  0,\pi/2\right)  .$ A
linear, bounded self-adjoint operator $A:\mathcal{H}\rightarrow\mathcal{H}$
with $\Delta_{w}A+\Delta_{v}A>0$ reaches Fleming's bound, i.e. obeys%
\begin{equation}
\delta_{A}:=\frac{\left\vert \left\langle A\right\rangle _{w}-\left\langle
A\right\rangle _{v}\right\vert }{\Delta_{w}A+\Delta_{v}A}=\tan\theta,
\end{equation}
if and only if

(i) $A$ stabilizes the subspace $\mathcal{H}_{0}=\mathbb{C}\cdot
v+\mathbb{C}\cdot w\subset\mathcal{H}$ and

(ii) the restriction of $A$ to $\mathcal{H}_{0}$ is related to an operator
$A_{0}$ from the set%
\begin{equation}
\left\{  \cos\left(  \alpha\right)  \left(  E_{11}-E_{22}\right)  +\sin\left(
\alpha\right)  \left(  E_{12}+E_{21}\right)  \left\vert \alpha\in\left[
\theta,\pi-\theta\right]  \right.  \right\}
\end{equation}
through $\left.  A\right\vert _{\mathcal{H}_{0}}=\lambda A_{0}+\mu\iota
d_{\mathcal{H}_{0}}$ for some $\lambda\in\mathbb{R}\setminus0$ and $\mu
\in\mathbb{R}.$ Here $E_{ij}:=e_{i}\left\langle e_{j},\cdot\right\rangle ,$
with the vectors $e_{i}$ from equation (\ref{ONB}).
\end{proposition}

Observe from equation (\ref{DiffExpt}) that for $\alpha=\pi/2$ the expectation
values' difference $\left\langle A\right\rangle _{w}-\left\langle
A\right\rangle _{v}$ is maximal. The maximal difference has the value
$2\sin\theta.$ From equation (\ref{DiffW}) we then obtain for $p_{1}%
=p_{2}=1/2$ that%
\begin{equation}
\sin\theta=2W^{A}\left(  D\right)  -1.
\end{equation}
Thus we have $W^{A}\left(  D\right)  =\frac{1}{2}\left(  1+\sin\theta\right)
, $ which coincides with the result of Jaeger and Shimony \cite{JS95} stated
in equation (\ref{minerrmax}). The corresponding observable $A_{0}$ has the
following particularly simple form%
\begin{equation}
A_{0}=e_{1}\left\langle e_{2},\cdot\right\rangle +e_{2}\left\langle
e_{1},\cdot\right\rangle =\frac{w\left\langle w,\cdot\right\rangle
-v\left\langle v,\cdot\right\rangle }{\sin\theta}.
\end{equation}

\ack We thank Florian Fr\"owis and Markus Penz for several stimulating discussions.


\section*{References}
\begin{thebibliography}{9}                                                                                                %
\bibitem {Ch00}A Chefles, \textit{Quantum state discrimination,} Contemp Phys
\textbf{41} (2000) 401 - 24

\bibitem {JS95}G Jaeger, A Shimony, \textit{Optimal distinction between two
non-orthogonal quantum states,} Phys Lett \textbf{A 197} (1995) 83-7

\bibitem {B01}S Barnett, \textit{Quantum limited state discrimination,}
Fortschr Phys \textbf{49} (2001) 909-13

\bibitem {F01}G N Fleming, \textit{Uses of a quantum master inequality,} Arxiv
preprint physics/0106077, 2001 - arxiv.org

\bibitem {F07}G N Fleming, \textit{Correlation coefficients and
Robertson-Schr\"{o}dinger uncertainty relations,} qaunt-ph/0703226v1, march 2007

\bibitem {Os10}L Ostermann, \textit{Fleming's quantum-master-inequality in
spin-1/2-systems,} Diplomarbeit Universit\"{a}t Innsbruck, 2010
\end{thebibliography}
\end{document}